\pdfoutput=1

\documentclass[pra,onecolumn,showpacs,amsmath,amssymb,longbibliography]{revtex4-1} 

\usepackage{amssymb}
\usepackage{amsmath}
\usepackage{graphicx}
\usepackage{color}
\usepackage{siunitx}
\usepackage{grffile}

\graphicspath{{Figures/},.}        

\newcommand{\imu}{\mathrm{i}}
\newcommand{\dint}[2][]{\!\mathrm{d}^{#1}#2\,}                 
\newcommand{\dfint}[3][]{\! \frac{\mathrm{d}^{#1}#2}{#3}\,}
\renewcommand{\Re}{\mathrm{Re}\,}
\renewcommand{\Im}{\mathrm{Im}\,}
\newcommand{\nooc}[1]{\sqrt{\varepsilon_{{#1}}}\frac{\omega}{c}}

\newcommand{\PF}{\Phi}  
\newcommand{\pf}{\phi}  

\graphicspath{{Figures/},.}         

\begin{document}

\title{Replacement of ensemble averaging by the use of a broadband source in scattering of light from a one-dimensional randomly rough interface between two dielectric media}

\author{Alexei A. Maradudin}
\affiliation{Department of Physics and Astronomy, University of California, Irvine CA 92697, U.S.A.}
\email{aamaradu@uci.edu}

\author{Ingve Simonsen}
\affiliation{Surface du Verre et Interfaces, UMR 125 CNRS/Saint-Gobain, F-93303 Aubervilliers, France}
\affiliation{Department of Physics, NTNU --- Norwegian University of Science and Technology, NO-7491 Trondheim, Norway}
\email{Ingve.Simonsen@ntnu.no}

\begin{abstract}
  By the use of phase perturbation theory we show that if a single realization of a one-dimensional randomly rough interface between two dielectric media is illuminated at normal incidence from either medium by a broadband Gaussian beam, it produces a scattered field whose differential reflection coefficient closely matches the result produced by averaging the differential reflection coefficient produced by a monochromatic incident beam over the ensemble of realizations of the interface profile function. 
\end{abstract}

\maketitle


\section{Introduction}

In theoretical calculations of some property of monochromatic light scattered from, or transmitted through, a randomly rough surface, such as the angular or spatial dependence of its intensity, what is actually calculated is the average of that  property over the ensemble of realizations of the random surface profile.  This procedure averages over the speckles that would be produced if monochromatic light were scattered by or transmitted through a single realization of  the random surface, and produces a  smooth angular or spatial dependence of the property of interest.

In an experiment this property is measured for a single realization of the random surface.  If the surface is illuminated by a monochromatic source, the resulting speckles have to be averaged in some way to produce the kind of smooth curve that  ensemble averaging yields.  This can be done by rotating or dithering the sample.  However, in some cases moving  the surface is not an option.  In such cases one can exploit the fact that the speckle pattern  depends on the wavelength of the incident light~\cite{New-1} to average over the speckles by using a broadband (polychromatic) beam to illuminate the surface instead of a monochromatic one. In an earlier paper~\cite{1-new} it was demonstrated that illuminating one realization if a one-dimensional randomly rough perfectly conducting surface by an s-polarized broadband Gaussian beam produced an intensity profile of the scattered field that closely matched the one produced by averaging the intensity of the scattered field produced by a monochromatic incident beam over the ensemble of realizations of the random surface profile function.

In this paper we explore the replacement of ensemble  averaging by the  use of an incident broadband Gaussian beam in the more realistic case where the one dimensional rough interface between two dielectric media is illuminated at normal incidence from either medium and the differential reflection coefficient of the scattered light is sought.

\section{Scattering Theory}
\begin{figure}[tb]
   \centering
  \includegraphics[height=0.35\linewidth]{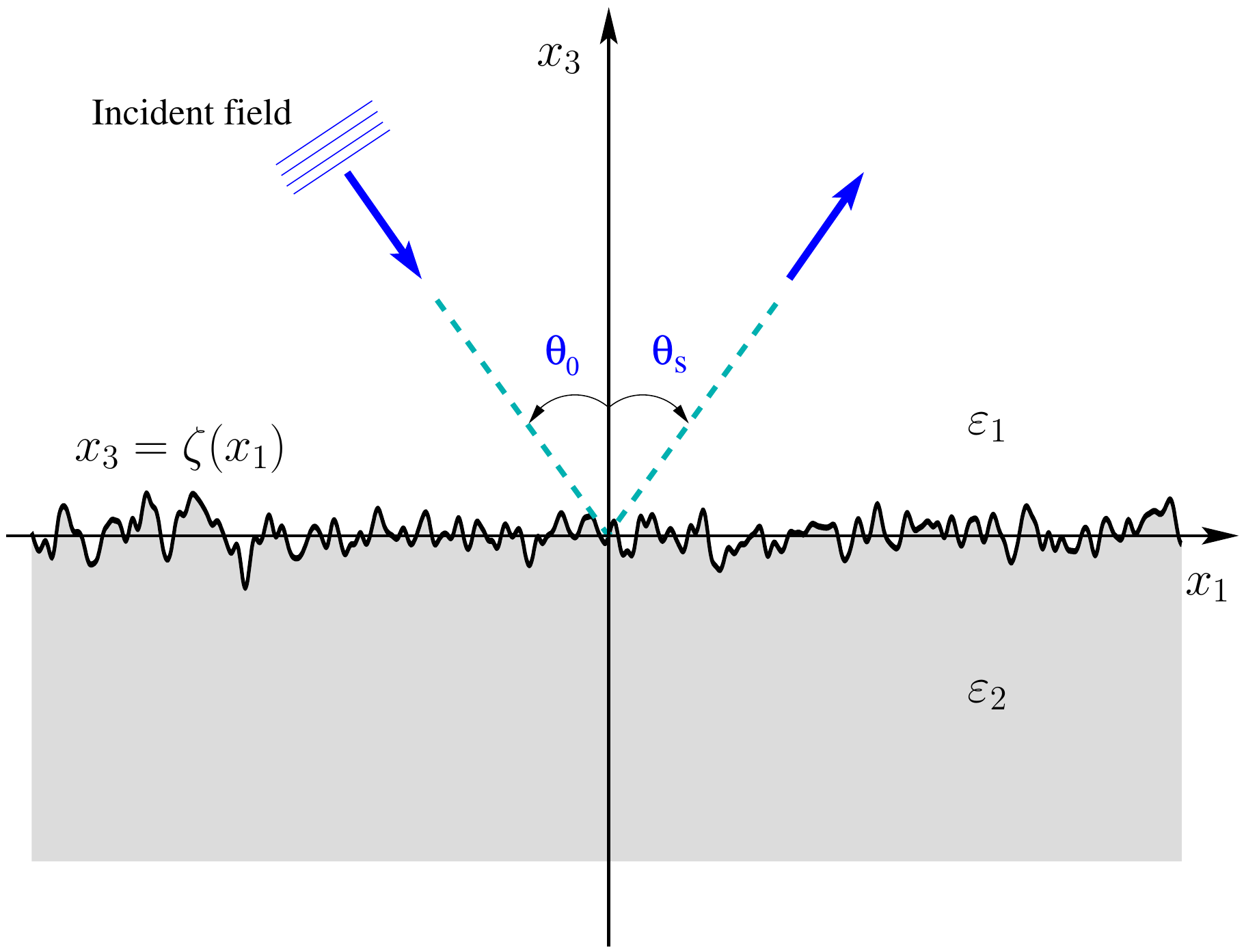}
  \caption{Schematics of the scattering geometry.}
     \label{Fig:0}
 \end{figure}

The system we study consists of a dielectric medium whose dielectric constant is $\varepsilon_1$ in the region $x_3 > \zeta(x_1)$, and a dielectric medium whose dielectric constant is $\varepsilon_2$ in the region  $x_3 < \zeta(x_1)$~[Fig.~\ref{Fig:0}].  Both $\varepsilon_1$ and $\varepsilon_2$ are assumed to be real, positive, and frequency independent.  The interface profile function $\zeta(x_1)$ is assumed to be a single-valued  function of $x_1$ that is differentiable and constitutes a random process.  This interface is illuminated at normal incidence from the region $x_3 > \zeta(x_1)$ by a p- or s-polarized  broadband Gaussian beam of light of angular frequency $\omega$, whose plane of incidence is the $x_1x_3$ plane.  The single nonzero component of its electromagnetic field is a weighted superposition of incoming plane waves,
\begin{align}
  {\mathcal F}_{\nu}(x_1,x_3;t)_{\textrm{inc}}
  &=
    \int\limits_{-\infty}^{\infty} \dfint[]{\omega}{2\pi}
    \int\limits_{-\nooc{1}}^{\nooc{1}} \dfint[]{k}{2\pi}
    W(k,\omega )
    \exp \left[\imu kx_1- \imu \alpha_1(k,\omega )x_3- \imu\omega t\right],
    \label{eq:1}
\end{align}
where ${\mathcal F}_{\nu}(x_1,x_3;t)_{\textrm{inc}}$ is $H_2(x_1,x_3;t)_{\textrm{inc}}$ when $\nu = p$, and is $E_2(x_1,x_3;t)$ when $\nu = s$.   The function $\alpha_1(k,\omega )$ is defined by $\alpha_1(k,\omega ) = [\varepsilon_1(\omega /c)^2 -k^2]^{\frac{1}{2}}$, with $\Re\alpha_1(k,\omega ) > 0$, $\Im \alpha_1(k,\omega ) > 0$, where $c$ is the speed of light in vacuum. The weight function $W(k,\omega )$ has the factored form
\begin{align}
  W(k,\omega )
  &=
    G(k)F(\omega ),
    \label{eq:2}
\end{align}
where
\begin{align}
  G(k)
  &=  
    \frac{2\sqrt{\pi}}{\alpha_1(k,\omega)}
    \sqrt{\varepsilon_1}\frac{w\omega}{2c}
    \exp\left[
    -\left( \sqrt{\varepsilon_1}\frac{w\omega}{2c} \right)^2
    \arcsin^2\left( \frac{k c}{\sqrt{\varepsilon_1}\omega} \right)
    \right],
    \label{eq:G_k}
\end{align}
while $F(\omega)$ is a random function that possesses the properties
\begin{subequations}
\label{eq:4}
\begin{align}
\langle F(\omega )F^*(\omega ')\rangle_{F} = & 2\pi \delta (\omega - \omega ') S_0(\omega )\label{eq:4a}\\
\langle F(\omega ) F(\omega ')\rangle_F = & 0 . \label{eq:4b}
\end{align}
\end{subequations}
The angle brackets $\langle \cdots\rangle_F$ here denote an average over the ensemble of realizations  of the field~\cite{1}.  An incident field of this nature is produced by a superluminescent diode~\cite{2}, for example.  We assume that the spectral density of the  incident field  $S_0(\omega )$ has a Gaussian form  with a central frequency $\omega_0$ and a $1/e$ halfwidth $\Delta \omega$,
\begin{align}
  S_0(\omega )
  &=
    \frac{1}{\sqrt{\pi}\Delta\omega}
    \exp  \left[- \left( \frac{\omega - \omega_0}{\Delta\omega } \right)^2 \right] .
    \label{eq:5}
\end{align}
In the following it will be assumed that the halfwidth is small enough that the spectral density of the incident light can be regarded as zero when $\omega<0$. Moreover, for convenience the function $F(\omega)$ will be regarded as zero when $\omega<0$.

%

In the limit that $\sqrt{\varepsilon_1} w\omega /2c \gg 1$, that will be assumed here,  Eq.~\eqref{eq:1} represents a Gaussian beam of $1/e$ half width  $w$ that is incident normally onto the rough interface. To see this, one make the change of variable $k=\sqrt{\varepsilon_1}(\omega/c)\sin\theta$ in Eq.~\eqref{eq:1} which results in a Gaussian integral on the right-hand-side of this equation that is evaluated analytically to produce  
\begin{align}
  {\mathcal F}_{\nu}(x_1,x_3;t)_\textrm{inc}
  &=
    \int\limits_{-\infty}^{\infty} \dfint[]{\omega}{2\pi }
    F(\omega )
    \exp
    \left[
    -\left(\frac{x_1}{w}\right)^2
    -\imu \sqrt{\varepsilon_1} \frac{\omega}{c}x_3 - \imu \omega t \right].
    \label{eq:6}
\end{align}

Due to the linearity of the scattering problem, the scattered field can be written as
\begin{subequations}
  \begin{align}
  {\mathcal F}_{\nu}(x_1,x_3;t)_{\textrm{sc}}
  &=
    \int\limits_{-\infty}^{\infty} \dfint[]{\omega}{2\pi}
    F(\omega)
    \int\limits_{-\nooc{1}}^{\nooc{1}} \dfint[]{k}{2\pi } G(k)
    \int\limits_{-\infty}^{\infty} \dfint[]{q}{2\pi}
    R_{\nu}(q|k)
    \exp
    \left[
    \imu qx_1
    + \imu \alpha_1(q,\omega ) x_3
    - \imu \omega t
    \right] ,
    \label{eq:7}
  \end{align}
  where $ R_{\nu}(q|k)$ is the scattering amplitude that is obtained when the incident field is given by $\exp[\imu kx_1-\imu\alpha_1(k,\omega)x_3-\imu\omega t]$,
or
  \begin{align}
  {\mathcal F}_{\nu}(r,\theta_s;t)_{\textrm{sc}}
  &=
    \int\limits_{-\infty}^{\infty} \dfint[]{\omega}{2\pi}
    F(\omega)
    \int\limits_{-\nooc{1}}^{\nooc{1}} \dfint[]{k}{2\pi } G(k)
    \int\limits_{-\infty}^{\infty} \dfint[]{q}{2\pi}
    R_{\nu}(q|k)
    \exp
    \left[
    \imu q r \sin\theta_s
    + \imu
    \alpha_1(q,\omega ) r \cos\theta_s
    - \imu \omega t
    \right],
  \end{align}
\end{subequations}
where $r=\sqrt{x_1^2+x_3^2}$ and $x_1=r\sin\theta_s$, $x_3=r\cos\theta_s$.
It can be calculated by any of several approaches such as small-amplitude perturbation theory~\cite{New-5}, the Kirchhoff approximation~\cite{New-6}, phase-perturbation theory~\cite{New-7}, and by rigorous numerical solutions of the equations of scattering theory~\cite{3}. We will use here the first-order phase perturbation theory expression for $R_\nu(q|k)$, due to its simplicity and because it iterpolates between small-amplitude perturbation theory and the Kirchhoff approximation.

In small-amplitude perturbation theory the expression obtained for $R_{\nu}(q|k)$ to the lowest nonzero order in the interface profile function $\zeta (x_1)$ is
\begin{align}
  R_{\nu}(q|k)
  &= R_{\nu}(k)
    \left[
    2\pi\delta(q-k)
    + \imu \PF_{\nu}(q|k) \hat{\zeta}(q-k)+\cdots
    \right] ,
    \label{eq:8}
\end{align}
where
\begin{align}
  \hat{\zeta}(Q)
  &=
    \int\limits_{-\infty}^{\infty} \dint{x_1}
    \zeta (x_1) \exp \left(-\imu Qx_1 \right) ,
    \label{eq:9}
\end{align}
while
\begin{subequations}
\label{eq:10}
\begin{align}
  R_p(k)
  &=
    \frac{
    \varepsilon_2 \alpha_1(k, \omega )
    -
    \varepsilon_1 \alpha_2(k, \omega )
    }{
    \varepsilon_2\alpha_1(k,\omega )
    +
    \varepsilon_1\alpha_2(k,\omega )
    }
    \label{eq:10a}
  \\
  \PF_p(q|k)
  &=
    \frac{
    \varepsilon_2-\varepsilon_1
    }{
    \varepsilon_2\alpha_1(q,\omega )
    +
    \varepsilon_1\alpha_2(q,\omega )
    }
    \left[
    \varepsilon_2qk
    -
    \varepsilon_1\alpha_2(q,\omega )\alpha_2 (k,\omega )
    \right]
    \frac{
    2\alpha_1 (k,\omega )
    }{
    \varepsilon_2\alpha_1(k,\omega )
    -
    \varepsilon_1\alpha_2(k,\omega )
    }
    \label{eq:10b}
\end{align}
\end{subequations}
and
\begin{subequations}
\label{eq:11}
\begin{align}
  R_s(k)
  &=
    \frac{
    \alpha_1(k,\omega ) - \alpha_2(k,\omega )
    }{
    \alpha_1(k,\omega ) + \alpha_2(k,\omega )
    }
    \label{eq:11a}
  \\
  \PF_s(q|k)
  &=
    \left[
    \alpha_2(q,\omega ) - \alpha_1 (q,\omega )
    \right]
    \frac{
    2\alpha_1 (k,\omega )
    }{
    \alpha_1(k,\omega )
    -
    \alpha_2(k,\omega )
    },
    \label{eq:11b}
\end{align}
\end{subequations}
with $\alpha_2(k,\omega ) = [\varepsilon_2(\omega /c)^2-k^2]^{\frac{1}{2}}$, $\Re\alpha_2(k,\omega ) > 0$, $\Im\alpha_2(k,\omega ) > 0$. We can rewrite the right-hand side of Eq.~\eqref{eq:8} as a Fourier integral,
\begin{align}
  R_{\nu}(q|k)
  &=
    R_{\nu}(k)
    \int\limits_{-\infty}^{\infty} \dint{x_1}
    \exp \left[-\imu(q-k)x_1 \right]
    \left[
    1 + \imu \PF_{\nu}(q|k)\zeta (x_1) + \cdots
    \right] . \label{eq:12}
\end{align}
On exponentiating the expression in brackets in the integrand in this expression, we obtain the first-order phase perturbation theory expression for $R_{\nu}(q|k)$,
\begin{align}
  R_{\nu}(q|k)
  &=
    R_{\nu}(k)
    \int\limits_{-\infty}^{\infty} \dint{x_1}
    \exp
    \left[
    -\imu (q-k)x_1
    \right]
    \exp
    \left[
    \imu \PF_{\nu}(q|k) \zeta(x_1)
    \right].
    \label{eq:13}
\end{align}
We will use this expression here due to its simplicity. After interchanging the order of the $k$ and $q$-integrations in Eq.~\eqref{eq:7} it follows that the scattering amplitude for a Gaussian monochromatic beam of frequency $\omega$ can be expressed as
\begin{align}
  R_\nu(q,\omega)
  &=
    \int\limits_{-\nooc{1}}^{\nooc{1}} \dfint[]{k}{2\pi}
    R_\nu(q|k) G(k),
    \label{eq:R_q_omega}
\end{align}
so that the scattered field becomes
\begin{align}
  {\mathcal F}_{\nu}(x_1,x_3;t)_{\textrm{sc}}
  &=
    \int\limits_{-\infty}^{\infty} \dfint[]{\omega}{2\pi}
    F(\omega)
    \int\limits_{-\infty}^{\infty} \dfint[]{q}{2\pi}
    R_{\nu}(q,\omega)
    \exp
    \left[
    \imu qx_1
    + \imu \alpha_1(q,\omega ) x_3
    - \imu \omega t
    \right].
    \label{eq:ScatteredField}
\end{align}
The scattering amplitude $R_\nu(q,\omega)$ enters the definition of the \textit{differential reflection coefficient} which is defined as the fraction of the power flux incident onto the rough interface that is scattered into an angular interval of width $d\theta_s$ about the scattering angle  $\theta_s$~\cite{New-10}. For an illumination of the random interface by a normally incident Gaussian beam of frequency $\omega$,  the expression for the differential reflection coefficient in the wide beam limit $\sqrt{\varepsilon_1} w\omega/(2c)\gg 1$ reads~\cite{3,Book:DesignerSurface}
\begin{align}
  \frac{\partial R_\nu}{\partial\theta_s}(\theta_s)
  &=
    \frac{\sqrt{\varepsilon_1}}{\sqrt{2}\pi^{3/2}} \frac{\omega}{cw}
    \cos^2\theta_s \left| R(q,\omega)\right|^2,
  \label{eq:drc}
\end{align}
where $q=\sqrt{\varepsilon_1}(\omega/c)\sin\theta_s$. When the differential reflection coefficient from Eq.~\eqref{eq:drc}, for monochromatic illumination at frequency $\omega_0$, is averaged over an ensemble of realizations of the random interface, the \textit{mean} differential reflection coefficient is obtained and we denote it $\left< \partial R_\nu/\partial \theta_s \right>$ in the following. On the other had, if the illumination of the random interface is done by a broadband source, characterized by the center frequency $\omega_0$ and the halfwidth $\Delta\omega$, the ``broadband'' differential reflection coefficient $\left< \partial R_\nu/\partial \theta_s \right>_F$  is obtained.     

We now turn to the calculation of a simple expression for the scattering amplitude $R_\nu(q,\omega)$. On substituting into Eq.~\eqref{eq:R_q_omega} the results from Eqs.~\eqref{eq:13} and \eqref{eq:G_k}, making the change of variable $k=\sqrt{\varepsilon_1}(\omega/c)\sin\theta$ in the resulting expression, one gets
\begin{align}
  R_\nu(q,\omega)
  &=
    \frac{\sqrt{\varepsilon_1}}{\pi} \frac{w\omega}{2c}
    \int\limits_{-\pi/2}^{\pi/2} \dint{\theta}
    \exp \left[-\left( \sqrt{\varepsilon_1} \frac{w\omega}{2c} \right)^2 \theta^2 \right]
    R_\nu\left(  \sqrt{\varepsilon_1}\frac{\omega}{c}\sin\theta \right)
    \nonumber
  \\
  &\qquad
    \times
    \int\limits_{-\infty}^\infty \dint{x_1}
    \exp
    \left[
    -\imu q x_1
    + \imu \sqrt{\varepsilon_1}\frac{\omega}{c}\sin\theta \,x_1
    \right]
    \exp
    \left[
    \imu \PF_\nu\left( q \big| \sqrt{\varepsilon_1}\frac{\omega}{c}\sin\theta \right) \zeta(x_1)
    \right].
    \label{eq:R_q_omega-tmp}
\end{align}
On passing to the limit $\sqrt{\varepsilon_1}(w\omega/2c) \gg 1$, corresponding to a wide Gaussian beam, one may take advantage of the approximations $\sin\theta\approx 0$ and $\cos\theta\approx 1$, with the consequence that $\PF_\nu$ becomes a \textit{linear} function of the variable $\theta$,
\begin{subequations}
  \label{eq:Phase-function-linear}
\begin{align}
  \PF_\nu\left( q \big| \sqrt{\varepsilon_1}\frac{\omega}{c}\sin\theta \right)
  &\approx
    \pf_\nu^{(0)}( q, \omega ) +  \pf_\nu^{(1)}( q, \omega ) \theta, 
\end{align}
with
\begin{align}
  \pf_\nu^{(0)}( q,\omega ) =  \PF_\nu( q | 0 )
\end{align}
and
\begin{align}
  \pf_p^{(1)}( q,\omega )
  &=
    2\frac{\omega}{c} \sqrt{ \varepsilon_1\varepsilon_2 }
    \left( \sqrt{\varepsilon_1} + \sqrt{\varepsilon_2} \right)
    \frac{q}{\varepsilon_2\alpha_1(q,\omega) + \varepsilon_1 \alpha_2(q,\omega) }
  \\
  \pf_s^{(1)}( q,\omega ) &= 0.
\end{align}
\end{subequations}
Moreover, in the limit $\sqrt{\varepsilon_1}(w\omega/2c) \gg 1$, the $\theta$-integral in Eq.~\eqref{eq:R_q_omega-tmp} takes the Gaussian form, due to the results in  Eq.~\eqref{eq:Phase-function-linear}, and can hence be evaluated analytically with the results that 
\begin{align}
R_\nu(q,\omega)
  &=
    R_\nu(0) 
    \int\limits_{-\infty}^\infty \dint{x_1}
    \exp
    \left[
    -
    \left\{
    \frac{x_1}{w}
    +
    \frac{ \pf^{(1)}_\nu(q,\omega) \zeta(x_1) }{ \sqrt{\varepsilon_1}\frac{\omega}{c}w }
    \right\}^2
    -
    \imu q x_1
    +
    \imu \pf_\nu^{(0)}(q,\omega) \zeta(x_1)
    \right].
  \label{eq:R_q_omega_final}
\end{align}
This is the simplified expression for the scattering amplitude derived from phase perturbation theory that we will use to produce the results to be presented later in this paper. It represents a significant simplification relative to, for instance, obtaining the scattering amplitude by rigorous means~\cite{3} which requires solving a linear system of equations that becomes time consuming when the interface becomes long.

\medskip
The interface profile function $\zeta(x_1)$ is assumed to be a single-valued function of $x_1$ that is differentiable and constitutes a zero-mean, stationary Gaussian random process defined by
\begin{align}
  \left< \zeta(x_1) \zeta(x_1')\right>
  &=
    \delta^2 W(|x_1-x_1'|).
\end{align}
The angles brackets here denote an average over the ensemble of realizations of $\zeta(x_1)$, $\delta=\left< \zeta^2(x_1)\right>^\frac{1}{2}$ is the root-mean-square height of the interface, and $W(|x_1|)$ is the normalized interface height auto-correlation function.

The power spectrum of the interface roughness, $g(|k|)$, is the Fourier transform of 
$W(|x_1|)$,
\begin{align}
  g(|k|)
  &=
    \int\limits_{-\infty}^{\infty} \dint{x_1}
    W(|x_1|) \exp\left( -\imu kx_1\right).
\end{align}
In the calculations carried out in this work, $W(|x_1|)$, will be assumed to have the Gaussian form
\begin{align}
  W(|x_1|)
  &=
     \exp\left( -\frac{x_1^2}{a^2} \right),
\end{align}
where the characteristic length $a$ is the transverse correlation length if the interface roughness. The power spectrum  $g(|k|)$ in this case also has the Gaussian form
\begin{align}
  g(|k|)
  &=
    \sqrt{\pi} a
       \exp\left( -\frac{k^2 a^2}{4} \right).
\end{align}

\medskip
A single realization of the interface profile function is given by~\cite{7}
\begin{align}
  \zeta(x_1)
  &=
    \delta \sqrt{\frac{2}{{\mathcal L}}}\,
    \sum\limits_{m=1}^{\infty}
    \left[
    g\left(\frac{2\pi m}{L} \right)
    \right]^\frac{1}{2}
    \left[
    \xi_{2m-1} \sin \left( \frac{2\pi m x_1}{{\mathcal L}}  \right)
    +
    \xi_{2m} \cos \left( \frac{2\pi m x_1}{{\mathcal L}}  \right)
    \right].
    \label{eq:random-surface-generation}
\end{align}
In this expression the $\{\xi_m\}$ are independent Gaussian random deviates with zero mean and unit variance:
\begin{align}
  \left< \xi_m\right> = 0,
  \qquad
  \left< \xi_m^2\right> = 1,
\end{align}
The function defined by Eq.~\eqref{eq:random-surface-generation} is a periodic function of $x_1$ with a period ${\mathcal L}$. To avoid edge effects only the portion of this function in the interval $-L/2<x_1<L/2$ where $L={\mathcal L}/2$, is used in calculations.

\section{Results and discussion}
\begin{figure}[tb]
   \centering
   \includegraphics[height=0.45\linewidth]{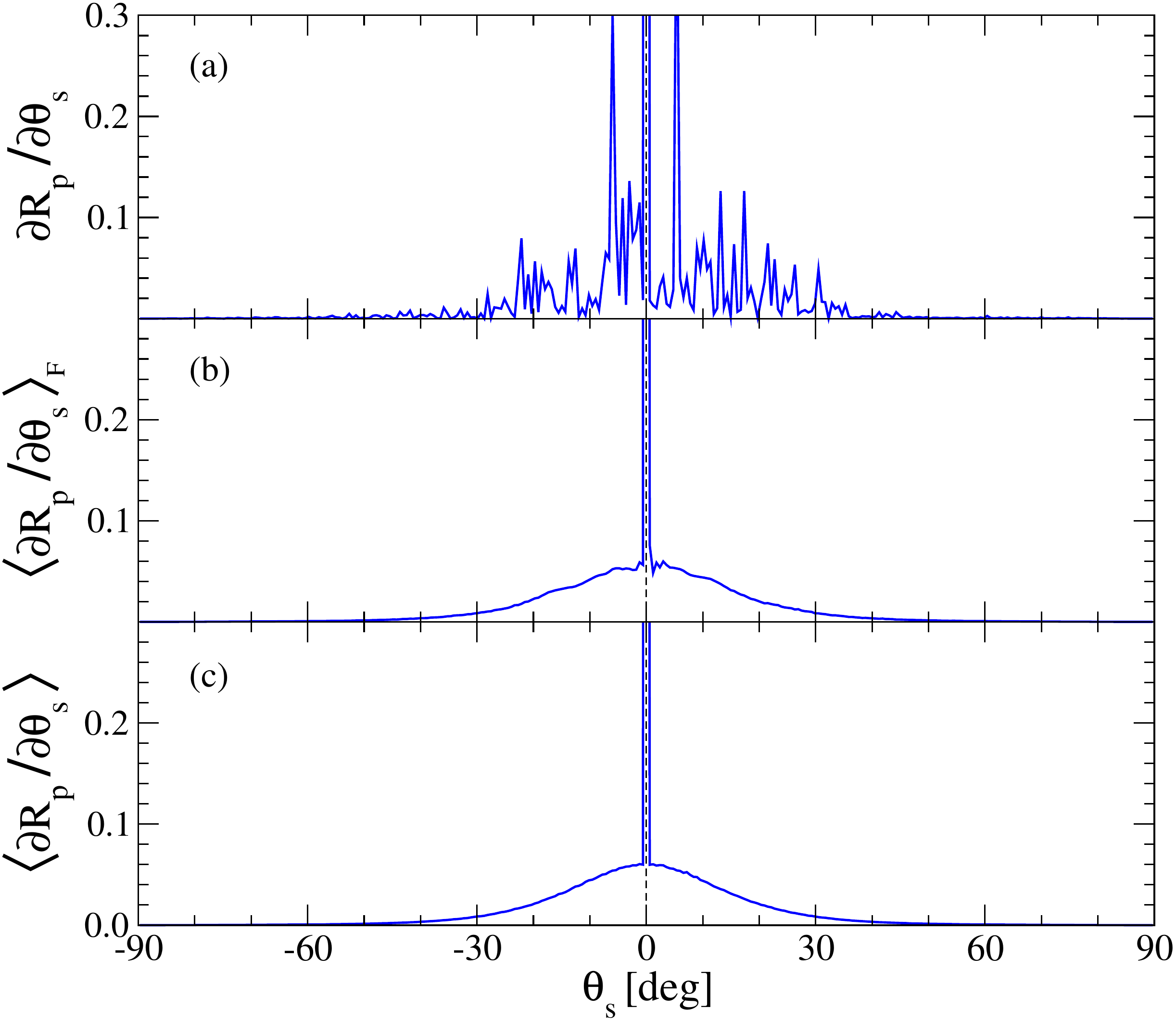}
   \caption{The differential reflection coefficients $\partial R_p/\partial\theta_s$ defined by Eq.~\eqref{eq:drc} for a vacuum-glass system [$\varepsilon_1=\num{1}$ and $\varepsilon_2=\num{2.25}$] illuminated from the vacuum side by different p-polarized incident beams: (a) a Gaussian monochromatic beam of halfwidth $w$ and frequency $\omega_0$; (b) a \textit{broadband} Gaussian beam of center frequency $\omega_0$ and frequency bandwidth $\Delta \omega=\num{0.2}\omega_0$; and (c) a monochromatic beam as in Fig.~\ref{Fig:1}(a) but with an average performed over an ensemble of  $N_p=\num{10000}$ realizations of the interface profile function. The angle of incidence was $\theta_0=\ang{0}$ and the wavelength $\lambda_0=2\pi c/\omega_0=\SI{632.8}{nm}$. The  randomly rough interface was characterized by the parameters
$\delta=\num{0.15}\lambda_0$, $a=\num{1.50}\lambda_0$, $L=\num{e4}\lambda_0$, and the sampling interval used to discretize the surface was $\Delta x_1=\lambda_0/10$. For the width of the Gaussian beam the value $w=L/4$ was used. The same random interface was used in producing the results of the first two panels of this figure.}
   \label{Fig:1}
 \end{figure}

\begin{figure}[tb]
   \centering
    \includegraphics[height=0.45\linewidth]{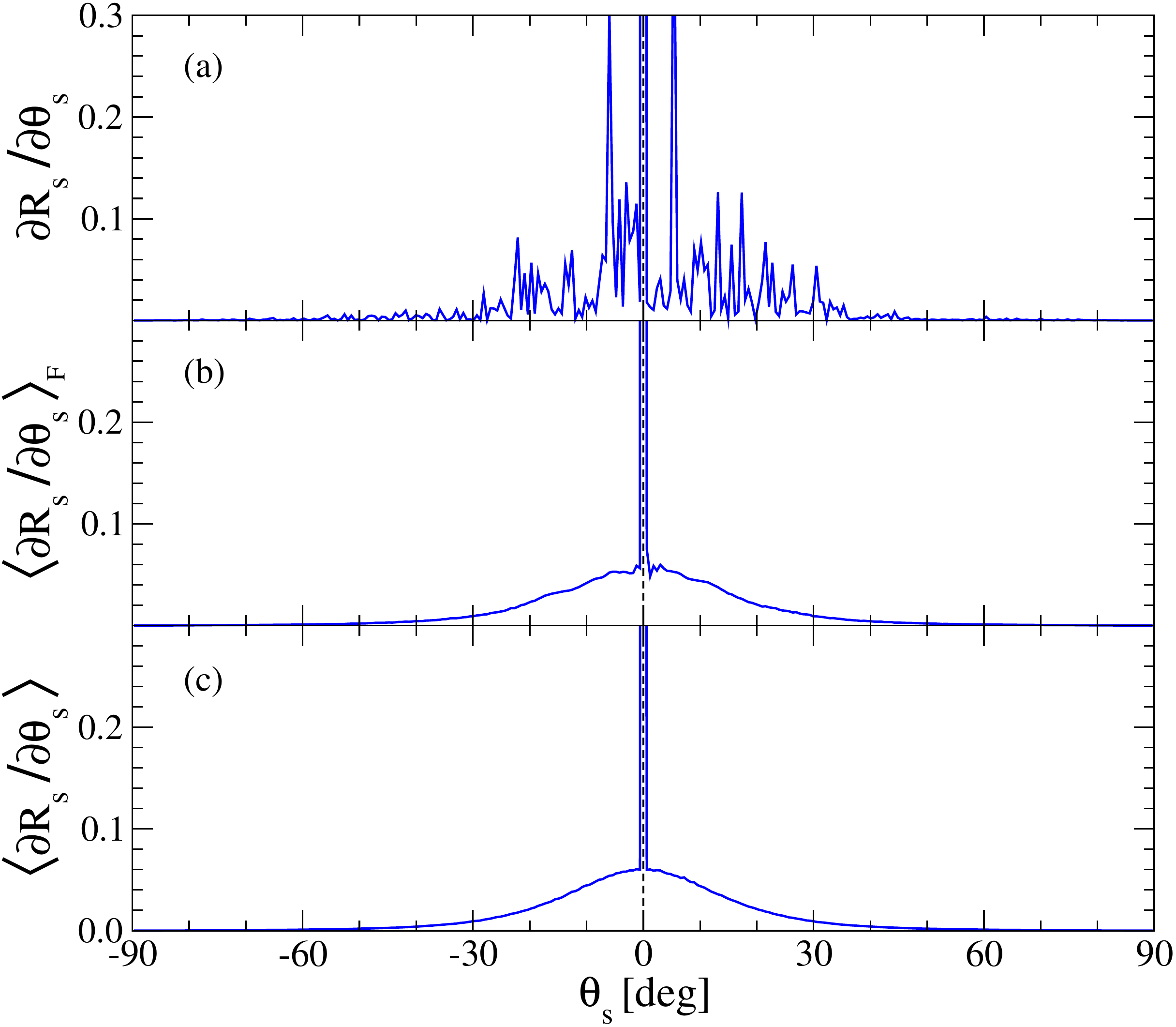}  
   \caption{Same as Fig.~\ref{Fig:1} but for s-polarized incident beams. The single realization of the rough interface used to obtain the results of the first two panels of this figure is the same one used in Fig.~\ref{Fig:1}.}
   \label{Fig:2}
 \end{figure}

In Fig.~\ref{Fig:1}(a) we present a plot of the speckle pattern of the scattered field as a function of the scattering angle $\theta_s$ produced by a monochromatic p-polarized Gaussian beam of frequency $\omega_0$, given by Eq.~\eqref{eq:6} with $F(\omega)=2\pi\delta(\omega - \omega_0 )$, incident on a single realization of a one-dimensional randomly rough interface generated with the use of  Eq~\eqref{eq:random-surface-generation}. In Fig.~\ref{Fig:1}(b) we present the differential reflection coefficient of the scattered field given by Eq.~\eqref{eq:ScatteredField} when the realization of the interface profile function used in obtaining Fig.~\ref{Fig:1}(a) is illuminated by a broadband Gaussian beam whose center frequency is $\omega_0$ with a halfwidth $\Delta\omega=0.2\omega_0$. To obtain this result calculations of the differential reflection coefficient were carried out for several values of $\Delta \omega$, ranging from $\num{0.2}\omega_0$ to $\num{0.4}\omega_0$. The results did not differ in any significant way so we chose to use the smallest of these $\Delta\omega$ values. It should be remarked that when a different realization of the surface was illuminated by the same broadband beam used to produce the result in Fig.~\ref{Fig:1}(b), the result was essentially $\left<\partial R_p/\partial\theta_s\right>_F$ from the same figure except for some small-amplitude fine details. Finally, in Fig.~\ref{Fig:1}(c) we plot the mean differential reflection coefficient $\left< \partial R_s/\partial\theta_s\right>$ obtained by averaging the results from $N_p=\num{10000}$ realizations of the interface profile function generated by Eq.\eqref{eq:random-surface-generation} when the interface is illuminated by the same monochromatic Gaussian beam of frequency $\omega_0$ used in obtaining Fig.~\ref{Fig:1}(a). The values of the theoretical and experimental parameters assumed in obtaining these results were $\varepsilon_1=1$,  $\varepsilon_2=2.25$, $\lambda_0=2\pi c/\omega_0=\SI{632.8}{nm}$, and $w=L/4$ where $L$ is the length of the surface. The parameters defining the interface roughness were $\delta=\num{0.15}\lambda_0$, $a=\num{1.50}\lambda_0$, and $L=\num{e4}\lambda_0$. The sampling interval used was $\Delta x_1=\lambda_0/10$ so that the interface was discretized onto $N=\num{e5}$ points. Figure~\ref{Fig:2} presents corresponding results for s-polarized incident beams. 

The results presented in Figs.~\ref{Fig:1} and \ref{Fig:2} show that the use of a broadband beam in illuminating a single realization of a one-dimensional randomly rough interface averages over the speckles produced by a monochromatic beam. It therefore produces a differential reflection coefficient that closely matches the one produced by a monochromatic beam when the resulting differential reflection coefficient is averaged over the ensemble of realizations of the interface profile function.

\begin{figure}[tb]
   \centering
   \includegraphics[height=0.5\columnwidth]{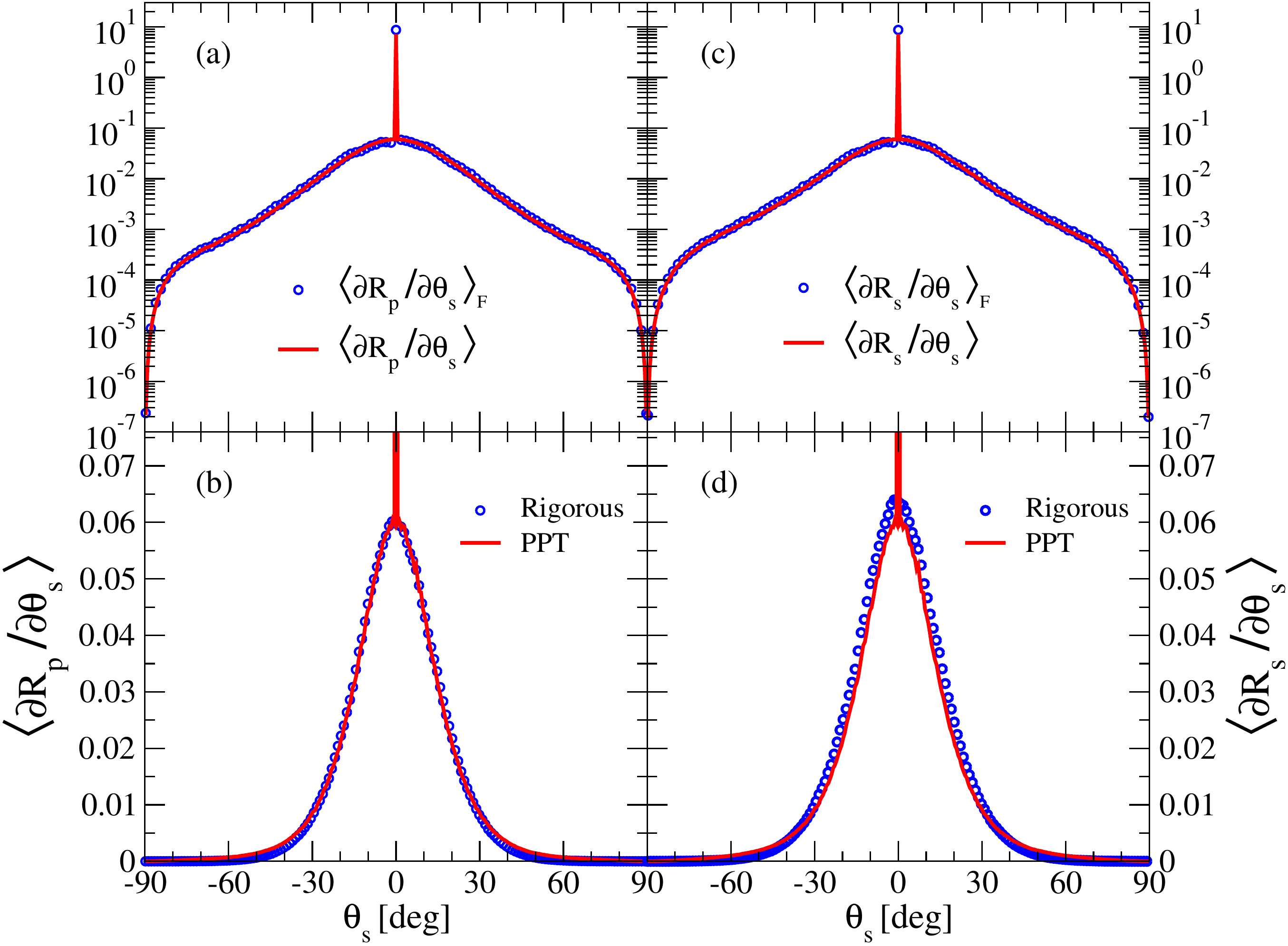}
   \caption{The quantities  $\left<\partial R_\nu/\partial\theta_s\right>_F$ and $\left<\partial R_\nu/\partial\theta_s\right>$ obtained for a vacuum-glass system [$\varepsilon_1=\num{1}$ and $\varepsilon_2=\num{2.25}$] plotted on semi-logarithmic scales for $\nu=p$~[Fig.~\ref{Fig:3}(a)] and $\nu=s$~[Fig.~\ref{Fig:3}(b)]. Comparison of the mean differential reflection coefficients $\left<\partial R_\nu/\partial\theta_s\right>$ obtained by either a rigorous simulation approach or on the basis of phase perturbation theory~(PPT) for p-polarized~[Fig.~\ref{Fig:3}(b)] or s-polarized~[Fig.~\ref{Fig:3}(d)] illumination. All ensemble averaged quantities were obtained on the basis of $N_p=\num{10000}$ realizations of the interface profile function. The remaining parameters are identical to those of Fig.~\ref{Fig:1}.}
   \label{Fig:3}
 \end{figure}

To facilitate the comparison of  $\left<\partial R_\nu/\partial\theta_s\right>_F$ and $\left<\partial R_\nu/\partial\theta_s\right>$, in Figs.~\ref{Fig:3}(a) and \ref{Fig:3}(c) we plot simultaneously these quantities on a semi-logarithmic scale. It is observed that the agreement between them is rather good even in the tails of the scattered intensity distributions.

The results presented in Figs.~\ref{Fig:1} and \ref{Fig:2} were, for convenience, all obtained under the assumption of phase perturbation theory. In Fig.~\ref{Fig:3} we compare the mean differential reflection coefficients $\left<\partial R_\nu/\partial\theta_s\right>$ from Figs.~\ref{Fig:1}(c) and \ref{Fig:2}(c)  to rigorous computer simulation results obtained by solving the equations of scattering theory~\cite{3}. The agreement between the two sets of results is rather convincing. In passing it should be noted that in performing the rigorous simulations the length used for the rough interface was  $L'=\num{e2}\lambda_0$ and the width of the Gaussian beam was $L'/4$. This length of the rough interface is two orders of magnitude shorter than the length used in obtaining the results based on phase perturbation theory.

\begin{figure}[tb]
   \centering
   \includegraphics[height=0.45\linewidth]{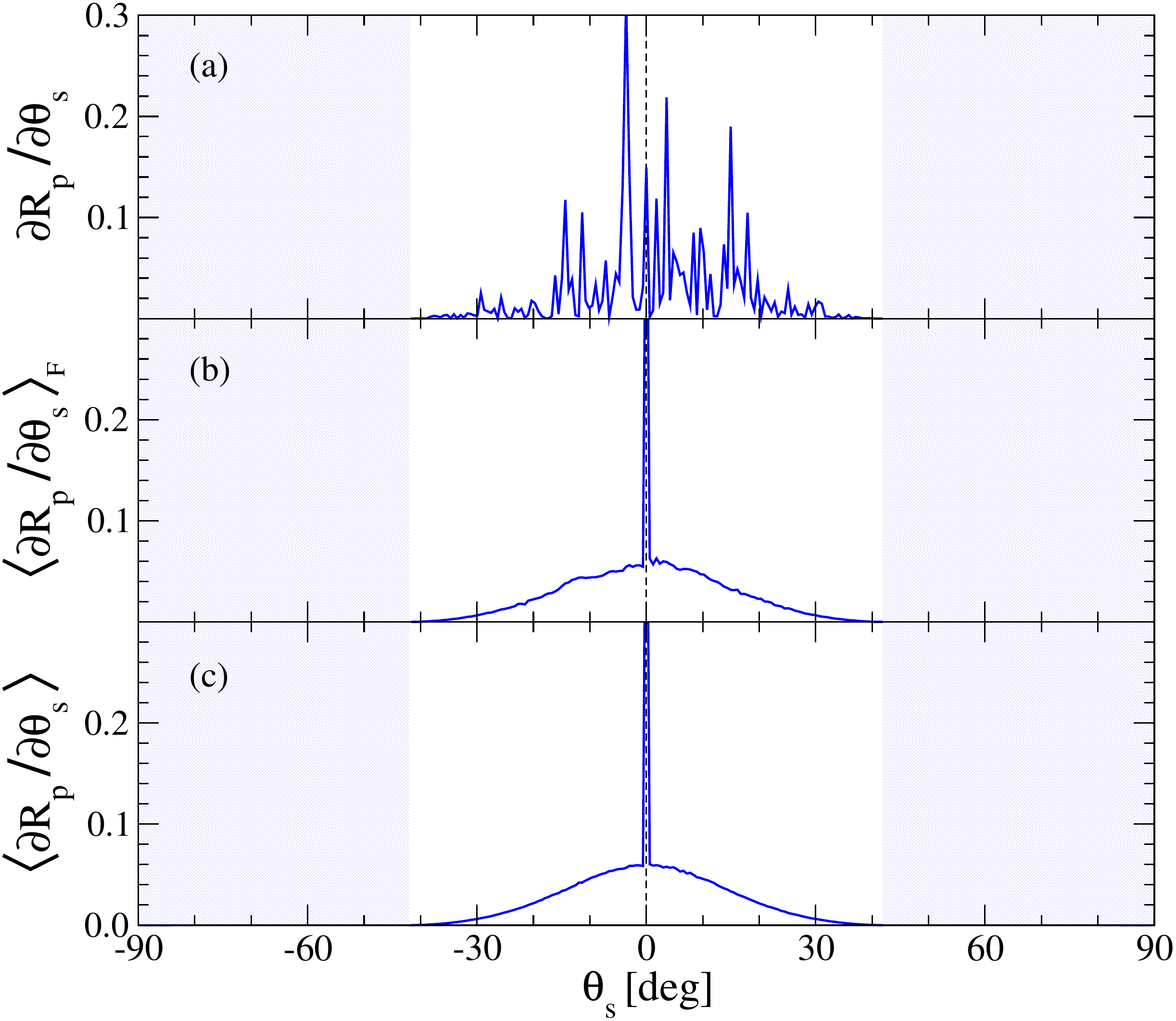}
   \caption{Same as Fig.~\ref{Fig:1}, \textit{i.e.} p-polarized illumination, but for a system where the medium of incidence is the optically denser medium [$\varepsilon_1=\num{2.25}$ and $\varepsilon_2=\num{1}$]. The shaded regions correspond to $|\theta_s|\geq \theta_s^\star$ where $\theta_s^\star=\arcsin(\sqrt{\varepsilon_2/\varepsilon_1})=\ang{41.81}$ denotes the critical angle for total internal reflection. The results are presented only for $|\theta_s|<\theta_s^\star$.}
   \label{Fig:4} 
 \end{figure}

\begin{figure}[tb]
   \centering
 \includegraphics[height=0.45\linewidth]{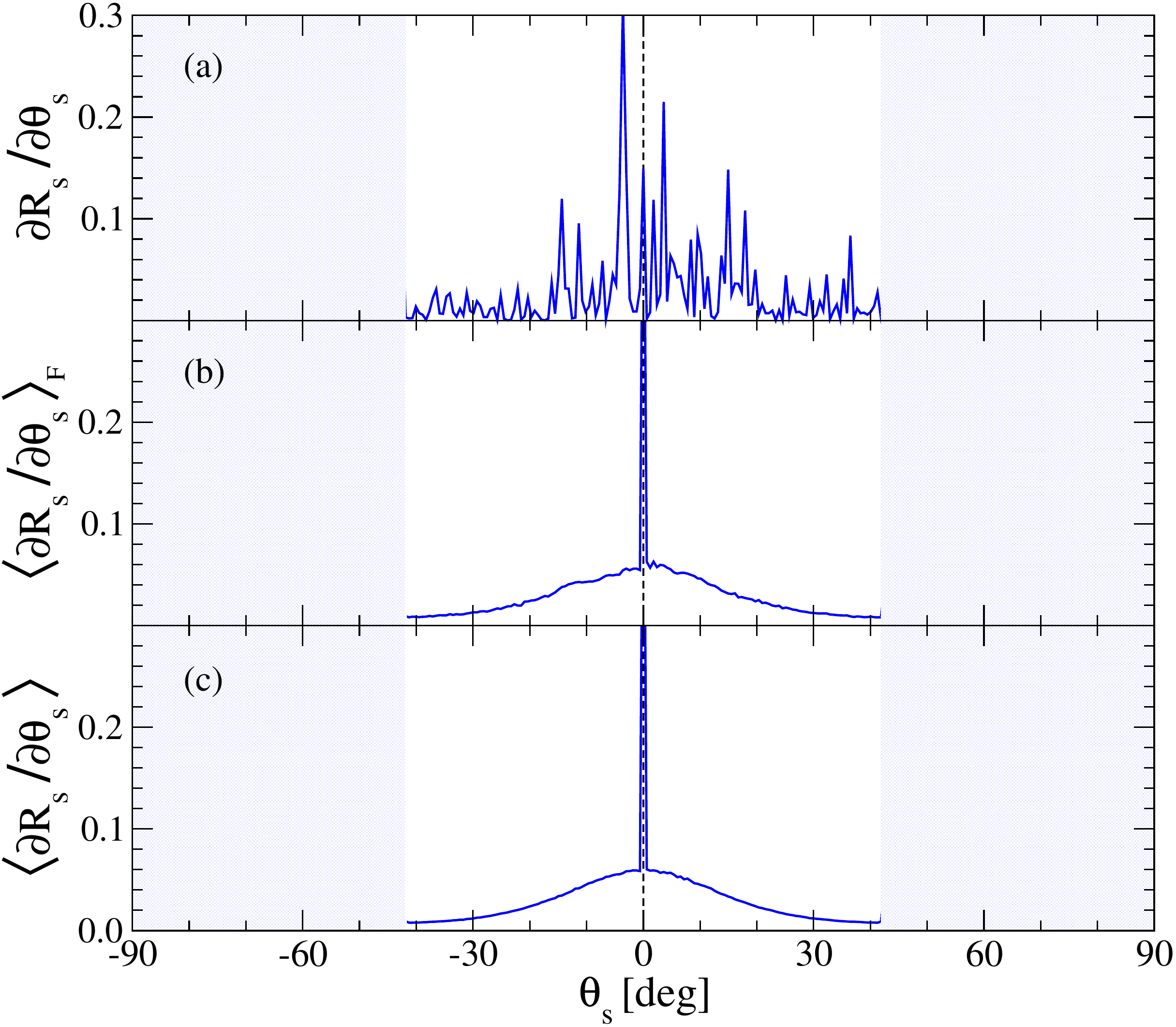} 
   \caption{Same as Fig.~\ref{Fig:4} [$\varepsilon_1=\num{2.25}$ and $\varepsilon_2=\num{1}$] but assuming s-polarized illumination.  
   }
   \label{Fig:5}
 \end{figure}

\begin{figure}[tb]
   \centering
   \includegraphics[height=0.5\columnwidth]{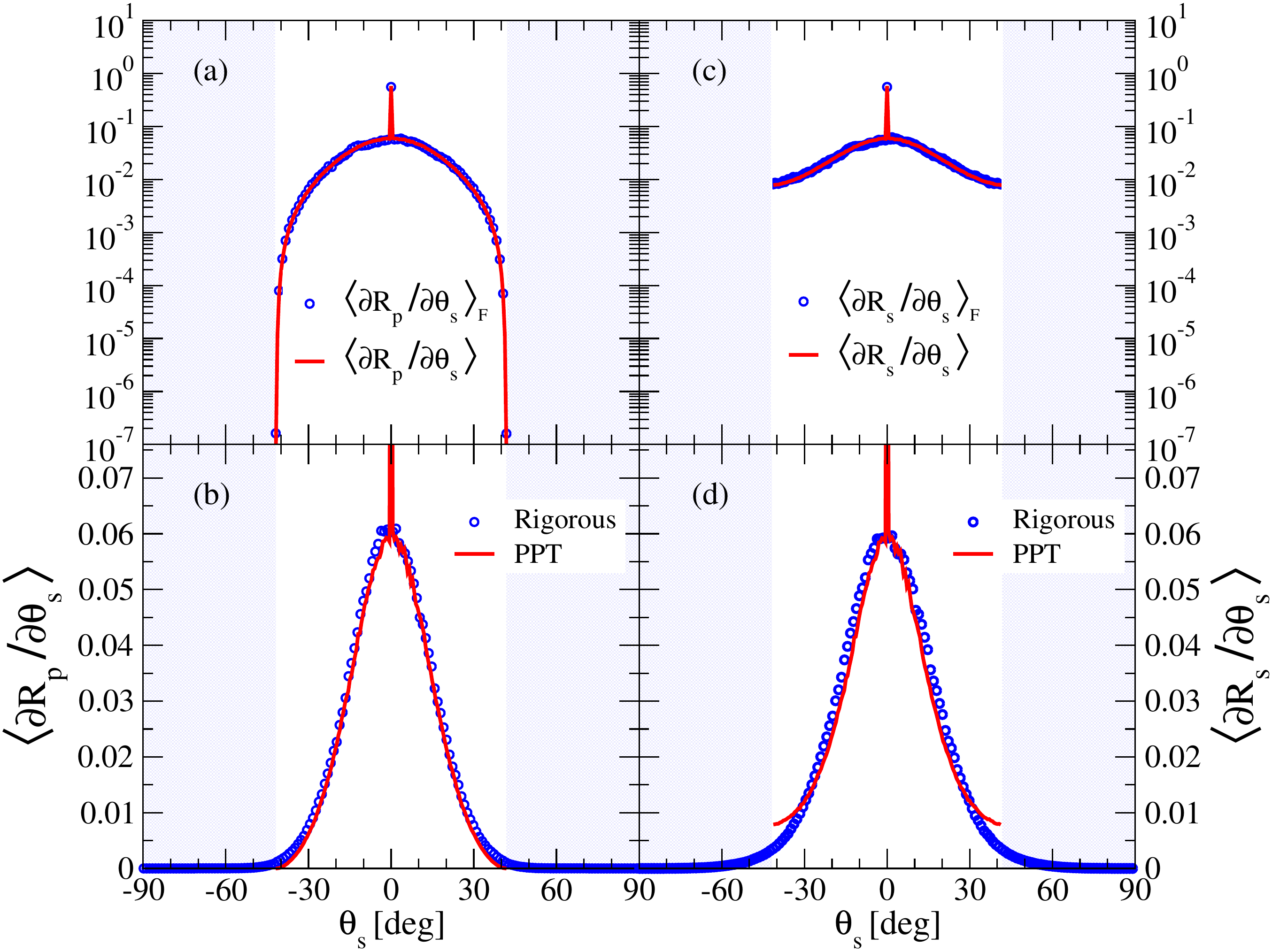}
   \caption{Same as Fig.~\ref{Fig:3} but for a glass-vacuum system [$\varepsilon_1=\num{2.25}$ and $\varepsilon_2=\num{1}$].}
   \label{Fig:6}
 \end{figure}

%
%
\medskip
We now turn to a scattering geometry where the medium of incidence is the optically denser medium. Here we assume a glass-vacuum system characterized by  $\varepsilon_1=\num{2.25}$ and $\varepsilon_2=\num{1}$. Physically this corresponds to the light being incident from the opposite side of the rough interface relative to the system we previously studied. The results for the differential reflection coefficients for the glass-vacuum system are presented in Figs.~\ref{Fig:4}--\ref{Fig:6}. In these figures, the angles for which  $|\theta_s| \geq \theta_s^\star$ have been indicated as shaded regions, where $\theta_s^\star=\arcsin(\sqrt{\varepsilon_2/\varepsilon_1})=\ang{41.81}$ denotes the critical angle for total internal reflection. Due to the assumptions underlying phase perturbation theory, it is expected not to work well when the angles of incidence and/or scattering in absolute value are larger than this critical angle. Hence, in Figs.~\ref{Fig:4}--\ref{Fig:6} we have only plotted the results obtained on the basis of phase perturbation theory for  $|\theta_s| < \theta_s^\star$.

On the basis of the results presented in Figs.~\ref{Fig:4}--\ref{Fig:6}, it is concluded that also for systems where the medium of incidence is the optically denser medium,  one finds  that the differential reflection coefficients $\left<\partial R_\nu/\partial\theta_s\right>_F$ and  $\left<\partial R_\nu/\partial\theta_s\right>$ match each other rather well in the angular interval $|\theta_s|<\theta_s^\star$.

%

%

%

%

%

%

\section{Conclusions}
The scattering of normally incident p- or s-polarized light from a one-dimensional randomly rough interface between two dielectric media is studied. Based on phase perturbation theory, it is demonstrated that using a broadband Gaussian beam to illuminate the surface  produces a differential reflection coefficient that closely matches the one produced by a monochromatic Gaussian beam when the resulting differential reflection coefficient is averaged over the ensemble of realizations of the interface profile function. This result is obtained since the broadband beam averages over the speckles produced by a monochromatic beam. 

The confirmation of the conjecture prompting the present work by these proof-of-concept calculations encourages additional calculations using rigorous computer simulation methods instead of the phase perturbation theory approach, to explore the efficacy of a broadband source in calculations of rough surface scattering phenomena and in experimental studies of them.

\section*{Conflict of Interest}
The authors declare that there is no conflict of interest regarding the publication of this paper.

\section*{Acknowledgments}
The research of I.S. was supported in part by the Research Council of Norway~(Contract 216699) and the French National Research Agency (ANR-15-CHIN-0003).


\end{document}